\documentclass[debug]{rmaa}

\usepackage{paralist}

\usepackage{psfrag,color}

\usepackage[latin1]{inputenc}

\title{The San Pedro M\'artir Kinematic Catalogue of Galactic Planetary Nebulae} 

\author{
  J. A. L\'opez,\altaffilmark{1} 
  M. G. Richer,\altaffilmark{1}
  M. T. Garc\'{\i}a-D\'{\i}az\altaffilmark{1}
  D. M. Clark,\altaffilmark{1} \\
  J. Meaburn, \altaffilmark{2}
  H. Riesgo, \altaffilmark{1}
  W. Steffen \altaffilmark{1} \& 
  M. Lloyd  \altaffilmark{2}
  }

\altaffiltext{1}{Instituto de Astronom\'{\i}a,
  UNAM, Campus Ensenada M\'exico.}

\altaffiltext{2}{Jodrell Bank Center for Astrohysics, University of Manchester,
  UK.}

\shortauthor{L\'opez et al.}
\shorttitle{The SPM Kinematic Catalogue of PN}

\fulladdresses{
\item J. A. L\'opez, M. G. Richer, M. T. Garc\'{\i}a-D\'{\i}az, D. M. Clark, H. Riesgo, W. Steffen: Instituto de Astronom\'{\i}a, Campus Ensenada, Universidad Nacional 
Aut\'onoma de M\'exico, Apartado Postal 877, Ensenada, Baja California, 22800, M\'exico
  (jal, richer, tere, dmclark, hriesgo, wsteffen @astrosen.unam.mx).
\item J. Meaburn and M. Lloyd: Jodrell Bank Center for Astrophysics,
  University of Manchester, Manchester, M13 9PL, UK (meaburnj@yahoo.co.uk, Myfanwy.Lloyd@manchester.ac.uk).}

\listofauthors{J. A. L\'opez, M. G. Richer, M. T. Garc\'{\i}a-D\'{\i}az, D. M. Clark, J. Meabrun, H. Riesgo, W. Steffen \& M. Lloyd}
\indexauthor{L\'opez, J. A.}
\indexauthor{Richer, M. G.}
\indexauthor{Garc\'{\i}a-D\'{\i}az, M. T.}
\indexauthor{Clark, D. M.}
\indexauthor{Meaburn, J.}
\indexauthor{Riesgo, H.}
\indexauthor{Steffen, W.}
\indexauthor{Lloyd, M.}

\abstract{
The San Pedro M\'artir kinematic catalogue of galactic planetary nebulae provides spatially resolved, long-slit, Echelle spectra for about 600  planetary nebulae. The data are presented wavelength calibrated and corrected for heliocentric motion. For most objects multiple spectra have been acquired and images with accurate slit positions on the nebula are also presented for each object. This is the most extensive and homogeneous single source of data concerning the internal kinematics of the ionized nebular material in planetary nebulae. Data can be retrieved for individual objects or selected by groups that share some common characteristic, such as by morphological classes, galactic population, binary cores, presence of fast outflows, etc. The catalogue is available through the world wide web at http://kincatpn.astrosen.unam.mx. 

}

\resumen{El cat\'alogo cinem\'atico de nebulosas planetarias gal\'acticas de San Pedro M\'artir provee espectros Echelle de rendija larga, resueltos espacialmente, para casi 600 nebulosas planetarias, represen. Los datos se presentan calibrados en longitud de onda y corregidos por movimiento helioc\'entrico. Para la mayor\'{\i}a de los objetos se han obtenido 
m\'ultiples espectros y para cada objeto se presentan im\'agenes  donde las posiciones de las rendijas han sido  indicadas con precisi\'on. Esta es la fuente individual m\'as grande y homog\'enea de datos relacionada con la cinem\'atica interna del material nebular ionizado en nebulosas planetarias. Los datos se pueden obtener para objetos individuales  o seleccionar grupos de objetos que comparten alguna caracter\'{\i}stica como es la clase morfol\'ogica, la poblaci\'on gal\'actica, n\'ucleos binarios, presencia de flujos de alta velocidad, etc. El cat\'alogo est\'a  disponible en la direcci\'on http://kincatpn.astrosen.unam.mx.}

\addkeyword{ISM: Planetary nebulae: general, catalogue}
\addkeyword{ISM: Jets and outflows}
\addkeyword{ISM: Kinematics and dynamics}
\addkeyword{Stars: Post-AGB}

\begin{document}
\maketitle

\section{Introduction}
\label{sec:intro}

The kinematics of the ionized nebular shell of planetary nebulae (PNe) provides key information on the physics that drives their expansion and which is necessary to understand their formation and evolution, as well as their role as galactic chemical contaminants of processed material through stellar mass-loss from the TP-AGB to the white dwarf stages. PNe are excellent tracers of galactic structure in our own and other galaxies and have been used as extra-galactic standard candles from the distribution of their [O III] luminosity function, which depends to a certain degree on the evolution of the ionized nebular shell. The complex morphological structures of PNe revealed in recent times by the {\it HST} and ground-based telescopes that achieve subarcsec image quality demand the knowledge of detailed expansion patterns to disentangle real outflows from projection, scattering and light-cone effects, particularly in the early stages of the ionized, forming nebular shells.

There have been a number of  compilations on the systemic heliocentric velocity of PNe. For example, \citet{Sh83} published heliocentric radial velocities for 524 PNe and \citet{DAZ98}  increased the data set to 867 PNe. These works have been useful for statistical studies and to derive the galactic rotation curve from the PNe population. However, the research on the internal motions in PNe started with the pioneering works of \citet{CM18}, \citet{Wi50} and \citet{OMW66}.  During the 1980's the field regained an intense activity (e.g. \citealp{RRA82, ChKKJ84}). \citet{Sa84} compiled expansion velocities for 165 PNe and \citet{We89} expanded this compilation from diverse sources to 288 PNe . Those works, together with many others on relatively small groups or individual objects (e.g. \citealp{MiSo90, LMP93, Lop98, GVL99, GZ00, GCM01, H07, Meab08, DVGM08}) have allowed to identify  differences in expansion patterns among different morphological classes and isolate components from different mass-loss episodes and age of the nebular shell. Furthermore, the growing high-quality data of the last decade on PNe has made clear the relevance of embedded collimated outflows, poly-polarity and other complex symmetries in their structural development (e.g. \citealp{SMV10, Ko10}). Unfortunately, detailed, spatially resolved kinematic information on the expansion patterns of the various components of the PNe shells can only be found for a relatively small number of objects scattered in the literature. Therefore, to understand the dynamics of PNe from an in-depth perspective a systematic and homogeneous set of high quality, spatially resolved, kinematic information of high spectral resolution, spanning most of the evolutionary stages, morphologies, progenitor masses and galactic populations is required. That is what the SPM kinematic catalogue on galactic planetary nebulae aims to provide. 

\section{The Observations and the Data}

The data for this first release of the SPM catalogue have been obtained over 55 observing runs. Fifty two observing runs were obtained at the 2.1 m, f/7.5 telescope in San Pedro M\'artir National Observatory, M\'exico, with the Manchester Echelle Spectrometer \citep{Meab03} and three additional runs in the southern hemisphere at the 3.9 m, f/8, Anglo-Australian Telescope, two of them using a twin MES and a third one with the UCLES spectrometer \citep{DI90} used in single order format. 

MES is an Echelle,  long-slit spectrometer optimized for nebular work. MES does not have a cross disperser, it isolates single orders using interference filters \citep{Meab03}. For example, the 87$^{th}$ order covers the H$\alpha$, \ion{He}{ii} $\lambda$ 6560 \AA,  \ion{C}{ii} $\lambda$ 6578 \AA, and the [\ion{N}{ii}] $\lambda$$\lambda$ 6548, 6583 \AA ~emission lines, and the 114$^{th}$ order contains the  [\ion{O}{iii}] $\lambda$ 5007 \AA~ line. The order containing the [\ion{S}{ii}] $\lambda$$\lambda$ 6717, 6731 \AA ~emission lines has also been observed for some objects. 
\begin{figure}[!t]
  \includegraphics[width=\columnwidth]{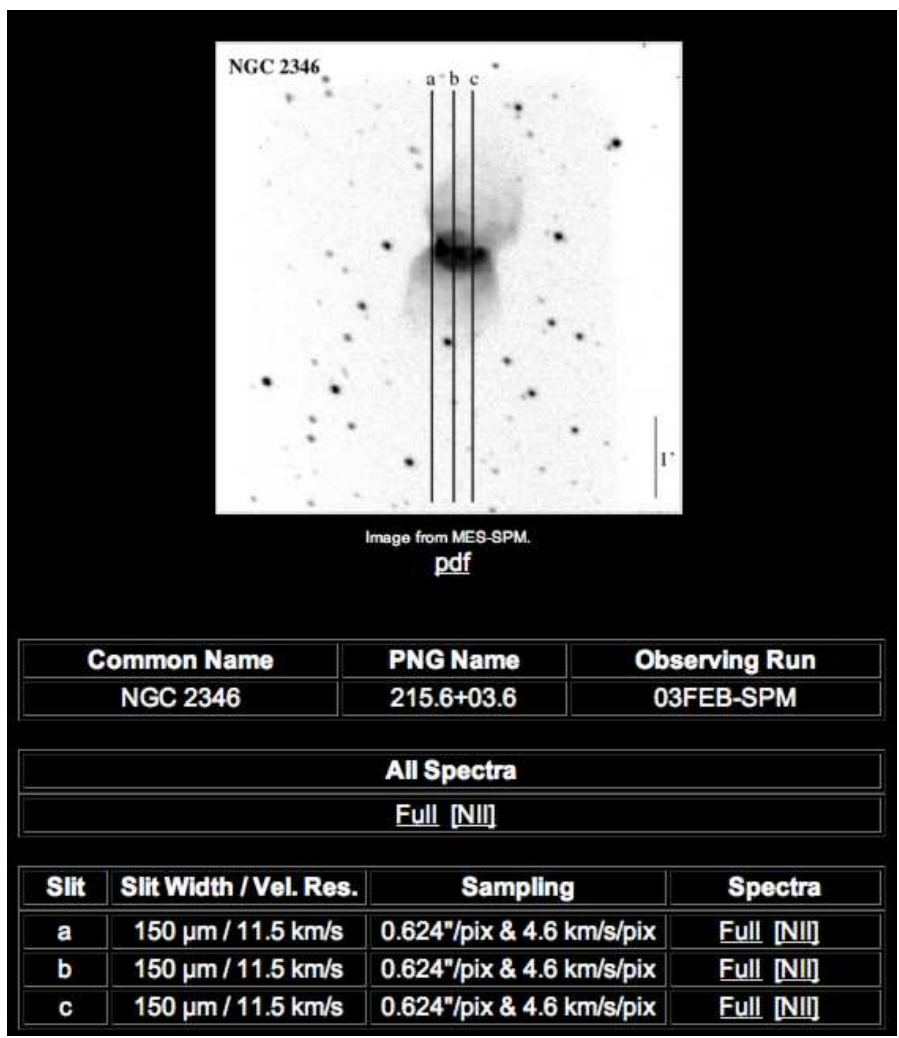}
  \caption{Example of the presentation of a PN in the catalogue.}
  \label{fig1}
\end{figure}

All of the spectra are contained in a database accessible via the world wide web at
 http://kincat.astrosen.unam.mx/. After selecting an individual PN an image of the nebula with accurate slit positions overlaid on it is presented and all the available data for that PN is displayed in a table below the image, see Figure 1.  The table provides at the top the common name and PNG identifier and the run(s) identifier for when it was observed. The data consists of  "Full" meaning the full spectral range (see Figure 2), calibrated in wavelength or individual bi-dimensional line spectra labeled by the corresponding line identifier such as [\ion{N}{ii}]  that refers to the $\lambda$ 6583 \AA ~emission line or [\ion{O}{iii}]  referring to the $\lambda$ 5007 \AA{} emission line (see Figure 3). In the latter case this is the only line of interest in the order, therefore this is never labeled as "Full" since it is always presented as an individual line spectrum. All individual line spectra are presented calibrated to heliocentric velocity. 

Next in the table and below the label \lq\lq All Spectra"  there are underlined keywords such as Full, H$\alpha$+[\ion{N}{ii}], [\ion{O}{iii}] or [\ion{N}{ii}], in case that the object has multiple observations. Clicking on any of these keywords all available data for the corresponding keyword for that object will be displayed to the right of the table. Below each image or spectrum there is a link for downloading a pdf file of it. Wavelength calibrated FITS files for individual objects can also be requested through the contact email address that appears in the front page of the web based catalogue. 
\begin{figure}[!t]
  \includegraphics[width=\columnwidth]{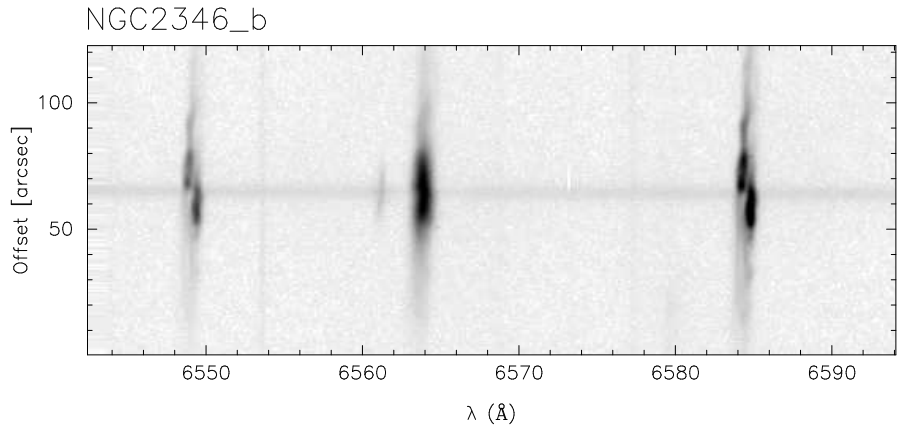}
  \caption{Full spectral range for the H$\alpha$ order corresponding to the central slit, b, in NGC 2346, see Figure 1.}
  \label{fig2}
\end{figure}

The full slit length in MES is in most cases 5\farcm2. This length is dependent on the CCD size format so it varies slightly in some cases depending on the CCD we used at the time. For the CCDs we used throughout these runs the slit varies from 5\farcm2{} to $\sim$ 7\farcm0.  Only in the case of the spectra obtained at the AAT with UCLES the slit length is  less than one arcmin, specifically 57\arcsec. In many cases it is not necessary to depict the full length of the slit on an image, only the portion that produces relevant data. All the images that accompany each object and show the location of the slits on them are oriented with north up and east left and show an angular scale bar. 
Likewise, all the position-velocity or bi-dimensional line profiles for each object are presented with an angular scale along the slit axis. Given the relatively large slit length most of the images are drawn either from the Digital Sky Survey or have been obtained with MES in its imaging mode, for only a few cases they are from other sources.  A legend at the bottom of the image with the slits shows the image origin. 
\begin{figure*}[!t]
  \includegraphics[width=0.45\linewidth,height=15cm]{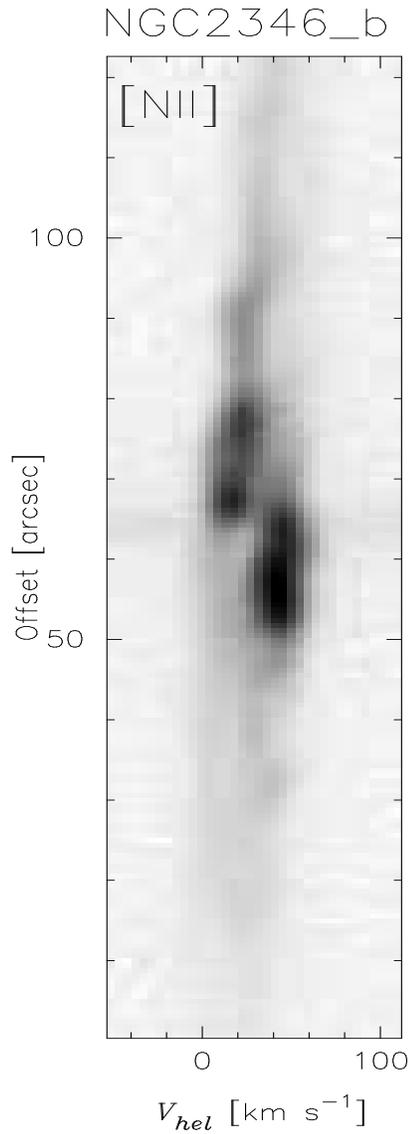}%
  \hfill
   \caption{Individual [\ion{N}{ii}] $\lambda$ 6583 \AA line profile corresponding to the central slit, b, in NGC 2346, see Figure 1.}
    \label{fig3}
\end{figure*}
The slit width has been for most objects 150 $\mu$m $\equiv$  11.5 km s$^{-1}$ in velocity resolution, for some bright objects we have used the 70 $\mu$m $\equiv$  6 km s$^{-1}$ and for some faint targets in some cases we have used a 300 $\mu$m $\equiv$  22 km s$^{-1}$. Since on some occasions we observed a target in more than one run, possibly with a different CCD or with different slit widths, or different on-chip binning, for each object we list in the associated table for that PN  all its associated slit positions, the slit width, its equivalent velocity resolution and associated spectral and spatial sampling information. Clicking on the keywords of the last column in the table, the individual \lq\lq Full" or line spectra are displayed.

\section{The database}

To connect to the database the user must provide her or his email address. After logging in the search page. The database provides the option to search for an individual PN by its common name or PNG name, following the IAU format. For the common name the prefix PN used by SIMBAD is excluded and for the PNG name the prefix PN G used by SIMBAD is also excluded.

In addition to searching for an individual object the entire database can be browsed  with the option Full List. The list can be ordered by common or PN G name. Objects can be searched and grouped by galactic coordinates or within a range of them. Additional, extended search methods are listed below that exploit the small wavelength range covered by the Echelle orders. For example, there is an option to search nebulae that show the presence or absence of the \ion{He}{ii} $\lambda$ 6560 \AA ~emission line, this serves as a method to search for high (evolved) or low (relatively younger) excitation objects. Likewise, there are options to search for objects in the database that lack the [\ion{N}{ii}] $\lambda$$\lambda$ 6548, 6583 \AA ~emission lines in their spectra. These PNe are usually high-excitation objects located near the end of the constant luminosity track and about to start the turn down in luminosity.
There is also an option to direct your search to select objects that show the presence of the only carbon recombination line, \ion{C}{ii} $\lambda$ 6578 \AA, located within the wavelength range covered by the H$\alpha$ order.

 During the course of putting together the catalogue we realized that there are a number of PNe that exhibit very wide H$\alpha$ wings, some are likely produced by Raman scattering of Ly $\beta$ photons into neighboring H$\alpha$ frequencies from high density nebulae, some symbiotic nebulae seem to be involved in this sample, others show P-Cygni profiles and others do show extended wings in H$\alpha$ and the  [\ion{N}{ii}] $\lambda$$\lambda$ 6548, 6583 \AA ~emission lines indicating true high-speed, low-emissivity, bipolar outflows. 
 
 The database allows also to retrieve those PNe that have been identified in different sources as containing a W-R or WELS type CS spectrum. Likewise, the catalogue also links PNe that are known to host binary and close binary central stars. PNe that exhibit high velocity outflows have also been provided a link, here the definition of high-velocity is usually applied, somewhat arbitrarily, for objects with outflows that show projected expansion velocities $\geq$ 70 km s$^{-1}$ and for objects where de-projected velocities are known to fulfill this criterion.  
 
 A coarse morphological classification has been applied to the nebula in the catalogue, describing just primary, main morphologies and descriptors. In some cases where the morphology was not clear from available images but it was clear enough from the basic geometry indicated by the line profiles and the radial velocity information, as described in \citet{SL06},   the classification was done based on the latter. If the spectra can neither provide a main morphological class then the nebulae were classified simply as compact.  In this way, PNe can be retrieved based on primary morphological class.   
 
 As it was to be expected, we have come across with some nebulae that are known to be or probably are mimics. We are providing a link in this case to identify them, comments from the community  that may help identify more of these cases from true PN are welcome. Additionally, it is also possible to select PNe based on their membership to a specific galactic population, as in the case with morphology, the classification is done only on primary descriptors, such as galactic location, systemic radial velocity and abundances in some instances, therefore revisions in some particular cases may be necessary. 
 
 Every time a group of PNe is selected and listed, the opportunity is also presented to the user to plot the group by galactic coordinates and to display the thumbnail images with their slit locations for all the members in that group.

Since the literature on the objects contained in the catalogue is so vast, we do not cite specific references for any of them, instead the catalogue provides a link at the top of the page for each PN to query SIMBAD automatically on that particular object if the user wishes to do so. Finally, at the end of the search page we have collected a number of related, useful www links. We welcome suggestions to include additional links in this section for the benefit of the user. 

 The ultimate objective is that this catalogue be an effective tool in helping us to better understand the formation and evolution of planetary nebulae. This work is a companion to the SPM planetary nebulae kinematic catalogue of extragalactic planetary nebulae \citep{Ri10} also available on line through one of the links mentioned above, at the end of the search page.

The authors gratefully acknowledge the numerous people that have provided valuable resources and contributed their time and talents to make this project possible. First we recognize the Committee for the Allocation of Telescope Time, CATT, for the San Pedro M\'artir National Observatory for the generous allocation of telescope time throughout the years. We acknowledge the skillful assistance of the telescope operators: Felipe Montalvo, Gabriel Garc\'{\i}a, Gustavo Melgoza and Salvador Monrroy and the whole technical staff of the Observatory that made possible to obtain data along 52 observing runs (about 300 nights) with less than 0.5\% down time due to technical failures. During the same runs we were only impeded to observe due to bad weather on 4 nights, so our thanks in this regard to whom it may concern up there. We thankfully recognize as well the Panel for Allocation of Telescope Time for the Anglo-Australian telescope where we obtained 3 different observing runs for southern targets. The Manchester Echelle Spectrometer (MES-SPM) resides in San Pedro M\'artir thanks to a collaborative agreement between  IA-UNAM and the University of Manchester. The participating members in this project, including students and postdocs, have benefited throughout the years from the generous financial support of DGAPA-UNAM through several PAPIIT projects as well as several CONACYT projects. Lastly, we are grateful to A. J. Clark for insightful discussions concerning the programming of the database.

\end{document}